\documentclass[sigconf,nonacm]{acmart} %

\AtBeginDocument{%
  }

\acmISBN{}

\setcopyright{none}

\usepackage{algorithm}
\usepackage{algorithmic}
\usepackage[inkscapelatex=false]{svg}
\usepackage[caption=false,font=footnotesize]{subfig}
\usepackage{graphicx}
\usepackage{float}
\usepackage{booktabs}
\usepackage{enumitem}
\usepackage{pifont}
\usepackage{array}

\begin{document}

\title{Aging-aware CPU Core Management for Embodied Carbon Amortization in Cloud LLM Inference}

\author{Tharindu B. Hewage\textsuperscript{1}, Shashikant Ilager\textsuperscript{2}, Maria Rodriguez Read\textsuperscript{1}, and Rajkumar Buyya\textsuperscript{1}}
\affiliation{%
  \institution{\textsuperscript{1}The University of Melbourne, Australia}
  \country{}
}
\affiliation{%
  \institution{\textsuperscript{2}University of Amsterdam, The Netherlands}
  \country{}
}

\renewcommand{\shortauthors}{}

\begin{abstract}
Broad adoption of Large Language Models (LLM) demands rapid expansions of cloud LLM inference clusters, leading to accumulation of embodied carbon$-$the emissions from manufacturing and supplying IT assets$-$that mostly concentrate on inference server CPU. This paper delves into the challenges of sustainable growth of cloud LLM inference, emphasizing extended amortization of CPU embodied over an increased lifespan. Given the reliability risks of silicon aging, we propose an aging-aware CPU core management technique to delay CPU aging effects, allowing the cluster operator to safely increase CPU life. Our technique exploits CPU underutilization patterns that we uncover in cloud LLM inference by halting aging in unused cores and even-outing aging in active cores via selective deep idling and aging-aware inference task allocation. Through extensive simulations using real-world Azure inference traces and an extended LLM cluster simulator from Microsoft, we show superior performance of our technique over existing methods with an estimated 37.67\% reduction in yearly embodied carbon emissions through p99 performance of managing CPU aging effects, a 77\% reduction in CPU underutilization, and less than 10\% impact to the inference service quality.

\end{abstract}

\keywords{Sustainable AI, Embodied Carbon Reduction, CPU Aging, LLM Inference}

\settopmatter{printfolios=true}
\maketitle

\section{Introduction}

\begin{figure}[t]
  \centering
  \includesvg[width=\columnwidth]{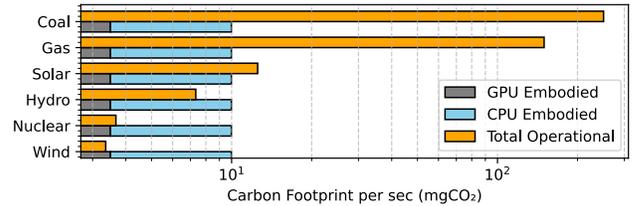}
  \caption{Carbon footprint of A100x4 GPU server running per second inference application when powered by energy sources with different carbon intensity \cite{li24carbon-efficient-llm}.}
  \Description{Carbon footprint of A100x4 GPU server running per second inference application when powered by energy sources with different carbon intensity \cite{li24carbon-efficient-llm}.}
  \label{fig:mt-em-vs-op-carbon}
\end{figure}

The proliferation of applications driven by cloud-based generative Large Language Model (LLM) inference is seen across diverse domains, such as conversational agents \cite{openai22intro-chat-gpt}, education \cite{bowne24edu-llm}, and coding assistance \cite{github21intro-copilot}. As the popularity of such applications scales their user base to billions \cite{fitimes24openai-1bn}, cloud service providers continue to expand LLM inference clusters towards the GigaWatt scale to support the growing demand \cite{crusoe24ai-dc-plans}. Recently, Meta announced its plans to build a brand new data center targeting AI workloads \cite{lagov24meta-new-ai-datacenter}, and xAI plans to expand their AI cluster from 100K to 1 million GPUs \cite{reuters24xai-memphis-ai-datacenter}.

LLM Inference clusters deploy and serve pre-trained LLMs \cite{patel24splitwise}. In inference, a user request (i.e. prompt query) is split into a set of input tokens. Input tokens are then fed to the model (i.e. forward pass) to generate the first output token and an intermediate context called KV-cache. KV-cache and the first token are then fed for the second forward pass, and the process is repeated until a stopping condition is met \cite{patel24splitwise}. In return, a single request can incur many forward passes. To reduce the latency in that, clusters employ parallel model computation via GPU accelerators \cite{rick23gpu-for-ai}. Due to memory constraints, each server typically utilizes several GPUs to support modern LLMs with billions of parameters \cite{schmid23llama}. In return, LLM inference becomes both memory and compute-intensive \cite{verma23llm-master}. When serving LLMs at scale, inference clusters employ many inference optimization techniques to better utilize underlying resources, such as phase splitting \cite{patel24splitwise} and iteration-level scheduling \cite{yu22orca}. As a result, LLM inference at scale results in a complex set of CPU tasks (i.e. inference tasks), such as facilitating steps of optimization techniques \cite{patel24splitwise,yu22orca}, request scheduling \cite{verma23llm-master}, and tokenization \cite{li24carbon-efficient-llm}.

 Growth of LLM inference clusters also increases cloud infrastructure carbon footprint \cite{bianchini2024dc-power-past-present-future, crusoe24ai-dc-plans, lagov24meta-new-ai-datacenter}. It combines two aspects: direct carbon emissions owing to energy sources (i.e. operational), and indirect carbon emissions results from business activities such as manufacturing, shipping and recycling IT assets (i.e. embodied) \cite{microsoft22role-of-embodied,zhang24emb-crb-models}. Today, as cloud service providers invest in renewable energy generation to meet net-zero emission goals \cite{crusoe24carbon-impact-llm,lagov24meta-new-ai-datacenter}, renewable energy sources with lesser carbon intensity \cite{maji24attribute-carbon} continue to penetrate power grids \cite{bianchini2024dc-power-past-present-future}, diminishing the effect of operational carbon over the embodied. Microsoft, a hyper-scale cloud provider reported that over the past four years, its operational carbon was reduced by 6.3 percent while embodied increased by 30.9 percent \cite{sally24ms-wood-dc}. In LLM inference clusters, most of its embodied carbon accounts for CPU components, including the die and mainboard \cite{li24carbon-efficient-llm}. Therefore, optimizing CPU embodied becomes paramount for sustainable growth of LLM inference clusters. Figure \ref{fig:mt-em-vs-op-carbon} illustrates that. With lesser carbon intensive renewable energy sources, CPU embodied becomes the dominant carbon aspect in inference servers.

 We study the problem of optimizing CPU embodied in LLM inference clusters. Inference clusters amortize CPU embodied over it's lifetime. Therefore, extending CPU life further amortize its embodied carbon. CPU life extensions are typically achieved through extending its hardware refresh cycle \cite{gupta22act}. CPU hardware refresh cycle replaces CPUs with newer hardware generations. Its aim is to gain performance-per-watt improvements \cite{gupta22act} and avoid reliability risks of silicon aging \cite{gnad15hyat,tiwari08facelift,gupta22act}. However, CPU performance gains in that are minimal for inference clusters. This is because CPU tasks in inference clusters carry-out GPU-accelerated LLM inference and these inference tasks mostly benefits from single core performance, which has plateaued in recent years \cite{tomlinson23extend-cpu-life-dc}. As a result, the sole aim from maintaining a standard hardware refresh cycle is to avoid silicon aging. In this context, extending CPU hardware refresh cycle requires efficient management of CPU to delay silicon aging effects. It is worth noting that this is not the case for GPU, for which the embodied carbon footprint is smaller and performance gains of newer hardware generation is significant \cite{li24carbon-efficient-llm}. Many works exploring silicon aging management in CPU employ efficient aging-aware workload management \cite{gnad15hyat, Shoulao23,zhao2023unsustainableaffinity,ansari23atlas}. They leverage task scheduling among CPU cores to even-out core aging and in return, slow down the aging rate of the overall CPU \cite{zhao2023unsustainableaffinity,tiwari08facelift,gnad15hyat,Shoulao23}. However, leveraging the opportunities present in CPU usage patterns of cloud LLM inference is yet to be explored. 

To this end, we propose an aging-aware CPU core management technique to extend the CPU life in inference clusters. In return, cluster embodied carbon is further amortized over the increased lifespan. We design our technique for CPU usage patterns in cloud llm inference. Using production inference traces, we uncover that LLM inference clusters mostly underutilize CPU cores with occasional usage bursts. To exploit that, we design a dynamic \textbf{working set} of cores where the cores in the set remain active while others deep idle \cite{intel18idle-time-mgt}. We then design online algorithms that (1) identify and adjust the \textit{working set} based on usage bursts, and (2) assign inference tasks inside the \textit{working set} to even-out aging across cores. Collectively, our approach achieves age-halting and reduced underutilization of cores. The \textit{working set} however, can lead inference tasks to oversubscribe the CPU if not scaled in-time. The online algorithms we propose are also designed to mitigate that.

We implement our approach by extending splitwise-sim, a high-fidelity LLM cluster simulator from Microsoft \cite{patel24splitwise}. We use production LLM inference traces generated with data collected from LLM inference services in Azure \cite{patel24splitwise} and use state-of-the-art CPU core management techniques as baselines. Results for our experimental cluster show; estimated 37.67\% reduction in yearly embodied carbon emissions through p99 performance of managing CPU aging effects and reduction of CPU core underutilization by 77\%, all while maintaining CPU oversubscription below 10\%. The \textbf{key contributions} of our work are as follows:
\begin{enumerate}[leftmargin=*, labelindent=0pt]
    \item An investigation into the role of the CPU in state-of-art LLM inference clusters and uncovering CPU underutilization patterns using production traces.
    \item A new technique for age-aware CPU core management using dynamic age-halting of deep idling CPU cores is proposed.
    \item Implement the proposed technique in a simulated environment and conduct extensive experiments using production inference traces.
    \item An evaluation of our proposed technique against state-of-the-art CPU core management baselines, focusing on its efficiency in managing CPU core aging, reducing yearly embodied carbon, and controlling task-related CPU oversubscription.
\end{enumerate}

\section{Background and Motivation}
\label{sec:background-and-motivation}

In this section, we provide background on embodied carbon amortization in cloud servers and optimizing it by extending CPU life. We then outline our investigations on applying that in cloud LLM inference clusters. We highlight key takeaways from where we draw our motivations for this paper.

\subsection{Background: Embodied Carbon Amortization in Cloud Servers through CPU Age Management}

In managing the carbon footprint of cloud data centers, Green House Gas (GHG) protocol, a global standard formed to manage GHG emissions \cite{ghg24protocol} defines three scopes. Scopes 1 and 2 represent the operational carbon, typically owing to the carbon intensity of the data center's energy sources. Scope 3 represents embodied carbon: carbon emissions that indirectly result from manufacturing and shipping of servers and other IT assets that have already been built and installed in data center \cite{microsoft22role-of-embodied}. Unlike operational carbon, which can be optimized by adopting less carbon-intensive energy sources, embodied carbon needs to be amortized over the asset's lifespan. Here, amortization is a way to account for embodied carbon. For example, if a server with a 4-year operational lifetime causes 1000 kgCO\textsubscript{2}eq of Scope 3 emissions, then amortization accounts for a 250 kgCO\textsubscript{2}eq of embodied carbon emissions per year.

Recent studies of embodied carbon optimization outline three tenets of environmental design: reduce, reuse, and recycle \cite{gupta22act}. Out of that, this paper focuses on recycling, more specifically enabling a second life of the CPU by improving its reliability to extend the lifetime \cite{gupta22act}. Primary reason for CPU reliability degradation is the silicon aging of its transistors beyond the rated life \cite{tiwari08facelift}. Many works studying silicon aging in CPU \cite{tiwari08facelift,gnad15hyat,saadatmand2021tamer} model that with \textbf{Negative Bias Temperature Instability (NBTI)}. NBTI is an aging mechanism that affects PMOS transistors in CPU \cite{ansari23atlas}. It is caused by the stress of workload execution. During workload execution, transistors in the CPU continue to switch, applying stress on transistors and releasing them back. When stress is applied, NBTI shifts the transistor's threshold voltage ($\Delta V_{th}$) but leaves a residual shift when the stress is removed. That incurs a slight increase in the $\Delta V_{th}$, accumulating over time. As a result, critical path delay in the circuit increases, reducing the maximum operating frequency of the CPU (i.e., \textbf{CPU Aging}). Since NBTI aging results from workload execution, workload management techniques can mitigate that for improved CPU reliability \cite{tiwari08facelift,gnad15hyat,saadatmand2021tamer, Shoulao23}. In return, embodied carbon is further amortized through a second life of the CPU.

\subsection{Motivation: Impact of CPUs’ on Embodied Carbon in LLM Inference Clusters}
\label{sec:motivation}

\begin{figure*}[htbp!]
  \centering
  \includesvg[width=0.85\linewidth]{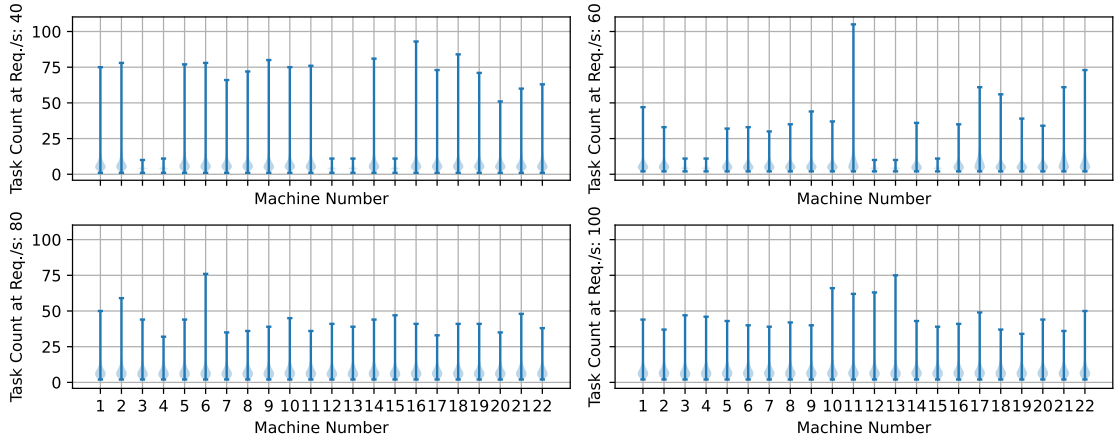}
  \caption{Distributions of running inference tasks in an LLM inference cluster of 22 H100 machines.}
  \Description{Distributions of running inference tasks in an LLM inference cluster of 22 H100 machines.}
  \label{fig:mt:cpu-util}
\end{figure*}

Servers in LLM inference clusters serve generative AI requests via low-latency model computation through GPU accelerators. Therefore, most of the inference server's thermal design power (TDP) comes from the GPU. In return, GPU dominates the server's operational carbon footprint \cite{li24carbon-efficient-llm}. In contrast, recent carbon modeling of inference servers shows CPU components such as the die and motherboard dominate the inference server's embodied carbon footprint \cite{li24carbon-efficient-llm}. In addition to that, the electrical grids powering inference clusters continue to integrate low-emission renewable energy sources \cite{iea23low-emmision-sources}. As a result, the operational carbon intensity of inference servers continue to diminish (see Figure \ref{fig:mt-em-vs-op-carbon}). In this context, the CPU embodied carbon becomes the dominant factor in LLM inference cluster's carbon footprint.

Inference clusters acquire CPU embodied faster than it can amortize through its hardware refresh life cycle. A hardware refresh cycle aims to gain performance improvements of newer CPU hardware generations and avoid reliability concerns of aging CPU \cite{gupta22act}. Recent studies show that LLM inference clusters may not gain significant performance benefits from newer CPU hardware generations. CPUs in inference clusters execute tasks (i.e., inference tasks) facilitating the inference workflows, such as phase splitting, request scheduling, batching, and tokenization \cite{patel24splitwise,li24carbon-efficient-llm}. These typically benefit from the single-core performance, yet the yearly single-core performance of newer CPU hardware generations has been mostly the same \cite{tomlinson23extend-cpu-life-dc}. That leaves avoiding reliability concerns of CPU aging as the sole benefit of the CPU hardware refresh cycle. As discussed in Section \ref{sec:background-and-motivation}, an efficient aging-aware workload management technique can mitigate CPU reliability concerns, which also translates to an extension to the CPU hardware refresh cycle, allowing the cluster to further amortize its CPU embodied. 

\noindent\textbf{Takeaways}: \textit{Extended amortization of CPU embodied significantly reduces the carbon footprint of inference clusters. However, it is constrained by the reliability concerns of CPU aging. Hence, there is an opportunity for an effective aging-aware CPU management technique to optimize that.}

\noindent\textbf{CPU Utilization Patterns in LLM Inference Clusters:} To design an efficient aging-aware CPU management technique, it is important to understand CPU utilization patterns in cloud LLM inference. For that, we monitor and analyze CPU utilization patterns in a high-fidelity simulated LLM inference cluster environment that infer real workload traces.

We use a LLM cluster simulator from Microsoft's \cite{patel24splitwise} and extend it to model CPU in cloud LLM inference. We then replay production inference traces from Azure and observe cluster CPU utilization. In that, we allocate each CPU task to a dedicated core. Our cluster settings closely match that in production. In Section \ref{sec::implementation},  We discuss our experimental setup in detail. Figure \ref{fig:mt:cpu-util} illustrates our results. Each subplot maps to a different throughput level and shows the distribution of concurrent inference tasks executed in each cluster machine. The x-axis denotes the machine number, and the y-axis denotes the inference task count. We make two key observations in that.

\begin{itemize}[leftmargin=*, labelindent=0pt]
    \item \textit{\textbf{O1:} Cores are mostly underutilized, as indicated by the lower mean values in the violin plots.}
    \item \textit{\textbf{O2:} There are occasional bursts of running tasks as indicated by the maximum values of the violin plots, which justifies having CPUs with higher core counts.}
\end{itemize}

In our simulation, our key finding is that underutilized CPU cores are available to the machine operating system to schedule system tasks apart from the inference serving platform. Therefore, these cores can actively execute system tasks in a time-shared manner \cite{yahya22agilewatts}. In return, all cores can actively execute instructions regardless of being allocated to an inference task, thus continuing to age due to the transistor stress of workload execution. In this context, we identify an opportunity to halt aging in the underutilized cores. We \textbf{\textit{hypothesize to reduce the available cores to match the number of running tasks of the inference platform}}. In return, we can put the remaining cores to deep idle, which turns off the clock and power gate the CPU cores \cite{yahya22darkgates,yahya22agilewatts}, stopping the transistor switching and halting the core aging. However, doing so can introduce a new set of challenges. As shown in Figure \ref{fig:mt:cpu-util}, the number of concurrent inference tasks can dynamically change. Therefore, limiting the available cores can lead inference tasks to oversubscribe the CPU unless the number of available cores is scaled in time. Moreover, a reduced set of available cores can introduce core affinity, which can increase failure risks of individual CPU cores due to uneven core aging \cite{zhao2023unsustainableaffinity}.

\noindent\textbf{Takeaways}: \textit{Underutilized cores in cloud LLM inference provide the opportunity to halt CPU aging through deep idling unused cores. However, it presents new challenges, including timely switching of core idle states to reduce CPU oversubscription, and efficient use of the available cores to avoid uneven core aging.}

\section{System Model and Problem Formulation}
\label{sec::system-model}

This section presents the system model, its components, and the formulation of the problem. 

\subsection{System Model}

\begin{figure}[t!]
  \centering
  \includesvg[width=0.92\columnwidth]{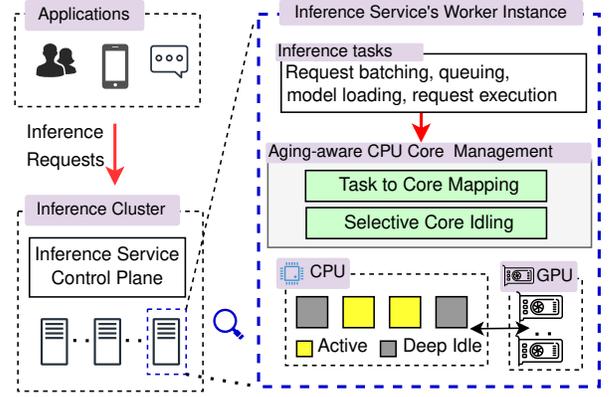}
  \caption{High-Level System Diagram of Aging-aware CPU Core Management in LLM Inference Clusters.}
    \Description{High-Level System Diagram of Aging-aware CPU Core Management in LLM Inference Clusters.}
  \label{fig:high-lvl-sys-diag}
\end{figure}

Figure \ref{fig:high-lvl-sys-diag} provides a high-level view of our system model. We model a high-performance LLM inference cluster deployment. Firstly, the inference requests from end-user applications reach the cluster's inference service. Each request is then scheduled to servers by the cluster-level scheduler. Inference servers run virtualized worker instances, which conduct request batching, queuing, model loading, and finally execute the request leveraging an inference backend, such as vLLM \cite{aws25llm-cluster-deploy}. Inference backend efficiently utilizes CPU and GPU resources and may leverage high-bandwidth InfiniBand interconnections between GPUs, such as sharing intermediate KV-cache in phase splitting \cite{patel24splitwise}. Our system architecture matches that of production deployments, such as the NVIDIA triton server inference architecture on Kubernetes \cite{aws25llm-cluster-deploy}. 

In return, the server-level worker instance of the inference service executes many CPU tasks (i.e., inference tasks). For each inference task, we allocate a dedicated CPU core. If the cores are insufficient, we assume that inference tasks oversubscribe the CPU. We introduce a new component to conduct aging-aware CPU core management. It oversees assigning cores to inference tasks and controlling the idle states of CPU cores. A CPU core in our system model can switch between either active or deep idle states \cite{intel18idle-time-mgt}. Being in the active state gradually ages the CPU cores. In contrast, deep idling halts cores from aging. However, cores that deep idle become unavailable for inference task execution. Cores in the active state are available for task execution, yet allocating a task accelerates core aging. In Section \ref{sec:aging-model}, we provide in-depth details about the silicon aging behavior with our aging model. CPU cores in the host are isolated and pinned to worker instance CPU cores, such that task mapping and idle state changes directly reflect on the physical core. Overall, the combination of inference task execution and deep idling control the CPU aging rate, where a lower value further amortizes its embodied carbon through the increased lifespan \cite{zhao2023unsustainableaffinity}.

\subsection{Aging Model}
\label{sec:aging-model}

We model aging of CPU cores due to execution of system and inference tasks. In that regard, Negative-bias Temperature Instability (NBTI) is a major aging mechanism \cite{gnad15hyat,ansari23atlas} for CPU. In this work, we model NBTI-induced core aging. Similar to previous works \cite{ansari23atlas}, we use a reaction-diffusion based aging model to calculate CPU core frequency degradation due to NBTI-induced aging.

\begin{equation}
    f(t) = f_0 \times (1 - \frac{\Delta V_{th}}{V_{dd} - V_{th}})
    \label{equation:core-freq}
\end{equation}
where $f(t)$ is the frequency at time $t$ and $f_0$ is the initial frequency of the core. Previous works show that $f_0$ can deviate from the nominal value due to variations in the manufacturing process \cite{raghunathan13cherry-picking, gnad15hyat}. To accommodate that, we use the following model to calculate $f_0$.

\[
f_0
= K' \,\min_{k,l \in S_{CP}}\!\Bigl(\tfrac{1}{p_{kl}}\Bigr).
\]
where $K'$ is a technology-dependent constant, and $S_{CP}$  represents the sections of the core containing critical paths. In order to calculate $f_0$, we first divide the chip area into an $N_{\text{chip}} \times  N_{\text{chip}}$ grid and assume that critical paths are contained entirely within the grid cells. Then, we assign each grid cell with a gaussian random variable ( $p_{kl}$). In order to calculate the spatial correlation between the random variables, we use the following formula \cite{raghunathan13cherry-picking}.
 
\[
\rho_{i j, k l} 
= e^{-\alpha \,\sqrt{(i - k)^2 + (j - l)^2}}
\quad \forall\, i, j, k, l \in [1, N_{\text{chip}}].
\]
where $\alpha$ decides how quickly spatial correlations die out. For our experiments, we set $N_{\text{chip}}$ to 10 and $K'$ to 1. Then, we set the mean of random variables by solving for a scenario where if a core does not exhibit process variation, $f_0$ should equal the nominal value. We set the remaining parameters similar to the calculation of a previous work \cite{raghunathan13cherry-picking}.

After $f_0$, we calculate $\Delta V_{th}$ in the Equation \ref{equation:core-freq}, which is the shift in the threshold voltage. CPU cores in our system can undergo time intervals in different idle states. In order to calculate $\Delta V_{th}$ in those, we calculate $\Delta V_{th}$ using the following recursive equation \cite{moghaddasi19nbti-estimate}.

\[
\Delta V_{\mathrm{th}}(t_p) 
= ADF_p \left[
    \left(\frac{\Delta V_{\mathrm{th}}(t_{p-1})}{ADF_p}\right)^{\frac{1}{n}}
    + \tau_p
\right]^n
\]
where $V_{\mathrm{th}}(t_p)$ is the value at $p^{th}$ time interval, and $\tau_p$ is the length of the $p^{th}$ time interval. $ADF$ is a time-independent factor for each time interval, which we calculate using the following equation.

\begin{figure}[t!]
  \centering
  \includesvg[width=0.9\linewidth]{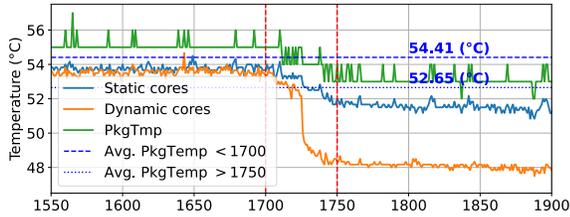}
  \caption{Changes in Operating temperature when 6 out of 12 cores set to deep idle in an Intel Xeon CPU. If awake, cores are 100\% utilized.}
    \Description{Changes in Operating temperature when 6 out of 12 cores set to deep idle in an Intel Xeon CPU. If awake, cores are 100\% utilized.}
  \label{fig:model:core-temp}
\end{figure}

\begin{equation}
ADF(T, V_{\mathrm{dd}}, Y)
= K \cdot \exp\!\Bigl(-\frac{E_{0}}{k_{B} T}\Bigr)
  \cdot \exp\!\Bigl(\frac{B\,V_{\mathrm{dd}}}{t_{\mathrm{ox}}\,k_{B} T}\Bigr)
  \cdot Y^{n}
\label{eq:adf}
\end{equation}
where $Y$ is the stress from the executing task. We assume each task in our system incur the worst case by setting it to 1.0. $T$ is the operating temperature of the CPU core. In order to create a realistic temperature model, we conduct an experiment by running a high utilization task in a server-grade CPU and switching its cores between active and deep idle states. During the experiment, we monitor the changes in core temperatures. Figure \ref{fig:model:core-temp} illustrates our observations. Table \ref{table:idle-temps} denotes the temperature model we derive from that. We set the rest of the parameters in equation \ref{eq:adf} as follows. $K$ is a fitting parameter. To calculate its value, we use CPU aging data from a previous work, which states that for 22nm CPU technology, the worst-case frequency reduction due to aging for a lifetime of 10-years can reach 30\% \cite{ansari23atlas}. We set values of our model to match this scenario and solve the $\Delta V_{th}$ equation to find the value for $K$. All parameters that were not explicitly mentioned are set similar to a previous work matching for the 22nm CPU technology \cite{ansari23atlas}.

\subsection{Problem Formulation}
\label{sec:prob-formulation}

This paper considers a cluster of LLM inference servers managed by an inference service. It serves inference requests arriving from cloud users.

At the server level, the GPU-accelerated inference requests results in a dynamic number of concurrent CPU tasks (i.e., inference tasks). They are handled by a multi-core CPU. Due to manufacturing process variations, each CPU core exhibits a maximum frequency value that deviates from the nominal value, degrading over time with workload execution due to core aging. Depending on the task allocation and management of core idle states, the rate of frequency degradation among the cores can differ. Over time, the multi-core CPU exhibits a distribution of degraded frequencies among its cores, increasing their failure risks \cite{zhao2023unsustainableaffinity}. In addition, the service quality of serving inference tasks can be impacted if active cores are insufficient to facilitate the running inference tasks. In this context, we formulate our problem as follows.

Our multi-core CPU contains an $N$ number of CPU cores. We denote the number of deep idling cores at time $t$ with $N_{idle}(t)$, the number of executing tasks with $T(t)$, and the frequency of the core $i$ with $f_{{core}_i}(t)$. For the duration of  $\Delta T$, the following equation calculates the reduction of core frequency due to core aging ($f_{{red}_i}(\Delta T)$) for the $i^{th}$ core.

\[
f_{{red}_i}(\Delta T) = f_{\mathrm{core}, i}(t) - f_{\mathrm{core}, i}(t + \Delta T)
\quad i \in N
\]

Similarly, the following equation calculates the variance in the CPU core frequency distribution ($f_{\mathrm{var}}(\Delta T)$).

\[
f_{\mathrm{var}}(\Delta T)
= \mathrm{Var}(F)
\quad \text{where }
F = \bigl\{\,f_{\mathrm{core}, i}\!\bigl(t + \Delta T\bigr)
: i \in N\bigr\}
\]

Finally, the following equation quantifies service quality impact from inference serving due to core deep idling ($T_{oversub}(\Delta T)$).

\[
\begin{aligned}
T_{oversub}(\Delta T) = \int_{t}^{t+\Delta T} u(T(t) - (N - N_{idle}(t))) \\ \times (T(t) - (N - N_{idle}(t)))
\end{aligned}
\]
Where $u$ is the unit step function. The goal of our problem is to extend amortizing CPU embodied carbon through increasing its operating lifespan by mitigating CPU aging effects. For that, both per-core and uneven aging effects across cores must be reduced. Moreover, the impact on the service quality of inference tasks must also be reduced. With that, we state our problem,
\[
\begin{aligned}
\mathrm{Min.} \quad  f_{{red}_i}(\Delta T) \quad \forall i \\
\mathrm{Min.} \quad f_{\mathrm{var}}(\Delta T) \\
\mathrm{Min.} \quad T_{oversub}(\Delta T)
\end{aligned}
\]

\begin{table}[tbp]
    \centering
    \caption{Temperature modelling for different core states.}
    \label{table:idle-temps}
    \begin{tabular}{>{\raggedright\arraybackslash}p{0.18\linewidth}>{\raggedright\arraybackslash}p{0.18\linewidth}>{\raggedright\arraybackslash}p{0.2\linewidth}>{\raggedright\arraybackslash}p{0.25\linewidth}} \hline 
          \textbf{Idle-state}&\textbf{C-state \cite{intel18idle-time-mgt}}&  \textbf{Inference Task}& \textbf{Temperature} ($^\circ C$)\\ \hline
          Active&C0&  Allocated& 54\\ 
          Active&C0&  Unallocated& 51.08\\  
          Deep Idle&C6&  N/A& 48\\ \hline
    \end{tabular}
\end{table}

\section{Extended Embodied Carbon Amortization through Aging-aware CPU Core Management in LLM Inference Clusters}

In this section, we provide the design and inner workings of our proposed aging-aware CPU core management technique. Our design is based on our findings in Section \ref{sec:motivation}. We evaluate the performance of the proposed technique in Section \ref{sec:perf-eval}, showcasing its potential to reduce yearly embodied carbon emissions of inference clusters.

Figure \ref{fig:high-lvl-sys-diag} illustrates the design of the proposed technique. We optimize CPU core aging at the server level to extend CPU lifetime, matching the CPU GPU asymmetric lifetime. To implement our proposed \textit{aging-aware core management}, we introduce two main mechanisms: \textit{(1) Task to Core Mapping}, and \textit{(2) Selective Core Idling}. \textit{Task to Core Mapping} runs an algorithm to decide the mapping of each inference task to a CPU core. It aims to mitigate uneven aging among available CPU cores. Whereas \textit{Selective Core Idling} runs an algorithm to determine a \textit{working set} of cores to match current inference throughput. It halts aging of cores in the non-working set by setting them to deep idle. Apart from age halting, it further complements uneven core aging by selecting cores to deep idle in an aging-aware manner.

\begin{algorithm}[htbp!]
\caption{Proposed Task-to-Core Mapping Algorithm.}
\label{algo:task-to-core-mapping}
\begin{algorithmic}[1]
\REQUIRE $cpu\_cores$: \textit{working set} of cores. Each has:
\begin{itemize}
  \item \texttt{task}: Inference task assigned (or \texttt{None} if no task),
  \item \texttt{last\_idle\_durations}: Recent idle durations.
\end{itemize}
\ENSURE The selected core to run the inference task.

\STATE $selected\_core \leftarrow \text{None}$ \label{algo:tsk2core:placeholders}
\STATE $selected\_idle\_score \leftarrow 0.0$

\FORALL{$core$ in $cpu\_cores$}
  \IF{$core.\text{task} \neq \text{None}$}
    \STATE \textbf{continue} 
  \ENDIF
  \STATE $idle\_score \leftarrow \sum(\text{core.last\_idle\_durations})$ \label{algo:tsk2core:idle-score}
  \IF{$(selected\_core = \text{None})$ \\ \OR $(idle\_score > selected\_idle\_score)$} \label{algo:tsk2core:comparison}
    \STATE $selected\_core \leftarrow core$
    \STATE $selected\_idle\_score \leftarrow idle\_score$
  \ENDIF
\ENDFOR
\RETURN $selected\_core$

\end{algorithmic}
\end{algorithm}

Together, the proposed technique reduces overall aging rate of the CPU in two aspects. Both \textit{Task-to-Core Mapping} and \textit{Selective Core Idling} \textbf{even-out aging} across cores, preventing early effects of premature aging in specific cores. \textit{Selective Core Idling} \textbf{halts aging} in cores, when the inference throughput provide opportunities to do so. It delays aging effects in cores.

\subsection{Task-to-Core Mapping}
\label{sec:sub:task2core}

The primary goal of the \textit{Task-to-Core Mapping} is to reduce the age variance among CPU cores. To achieve that, it distributes the stress of inference tasks favoring lesser-aged cores. As a result, older cores age slower, delaying overall aging effects.

Algorithm \ref{algo:task-to-core-mapping} outlines the proposed algorithm for \textit{Task-to-Core Mapping}. It takes the set of active cores (i.e., \textit{working set}) as the input and selects a core to run an inference task. Therefore, each new inference task executes the algorithm \ref{algo:task-to-core-mapping} once. To reduce the execution time in that, we design the algorithm to leverage an age estimation approach for its selection logic, rather than obtaining CPU micro-architectural attributes to calculate an accurate value. To achieve that, each core in the input \textit{working set} provides two additional attributes: task assigned status, and its idle history. Using a core's idle history, we calculate an estimation for it's age. We maintain a core's last eight idle durations, similar to that of the Linux governor algorithm \cite{intel18idle-time-mgt}. At execution, we create placeholders for both the selected core and its idle score (line \ref{algo:tsk2core:placeholders}). Then, we iteratively evaluate each core in the \textit{working set}. We calculate an idle score for each core that has not been assigned a task yet (line \ref{algo:tsk2core:idle-score}). Idle score accumulate all idle durations in the provided history. Here, the insight is that if a core mostly remained idle, its aging rate is lower than that of a less idle core. We then use the idle score to conduct a relative comparison among the cores to filter the core with the most idle score (line \ref{algo:tsk2core:comparison}). As a result, the core with the least aging estimation is selected to execute the next inference task.

\begin{algorithm}[htbp!]
\caption{Proposed Selective Core Idling Algorithm.}
\label{alg:adjust_sleeping_cores}
\begin{algorithmic}[1]
\REQUIRE 
\begin{itemize}
    \item \(\text{cpu\_cores}\): List of available cores
    \item \(\text{oversub\_tasks}\): Number of CPU oversubscribing tasks
\end{itemize}
\ENSURE Adjusted core idle states: deep idle or active.

\STATE \(N \leftarrow get\_total\_core\_count(\text{cpu\_cores})\) \label{algo:slctcoreidle:counts}
\STATE \(\text{active\_cores} \leftarrow get\_active\_core\_count(\text{cpu\_cores})\)
\STATE \(\text{normal\_tasks} \leftarrow get\_assigned\_task\_count(\text{cpu\_cores})\)
\STATE \(C_{SLP_t} \leftarrow N - \text{active\_cores}\) \label{algo:slctcoreidle:cslp}
\STATE \(T_t \leftarrow \text{normal\_tasks} + \text{oversub\_tasks}\)
\STATE \(T_t \leftarrow \min(N, T_t)\) \label{algo:slctcoreidle:capping}
\STATE \(e_t \leftarrow \bigl(N - C_{SLP_t} - T_t\bigr)\)
\STATE \(e_{t\_prd} \leftarrow e_t\)
\STATE \(e_{t\_prd} \leftarrow \frac{e_{t\_prd}}{N}\) \label{algo:slctcoreidle:error}
\IF{\(e_{t\_prd} \ge 0\)} \label{algo:slctcoreidle:react-start}
  \STATE \(F(e_{t\_prd}) \leftarrow \tan\bigl(0.785 \cdot e_{t\_prd}\bigr)\)
\ELSE
  \STATE \(F(e_{t\_prd}) \leftarrow \arctan\bigl(1.55 \cdot e_{t\_prd}\bigr)\)
\ENDIF \label{algo:slctcoreidle:react-end}
\STATE \(e_{t\_corr} \leftarrow N \times F(e_{t\_prd})\) \label{algo:slctcoreidle:scale-back}
\STATE \(e_{t\_corr} \leftarrow \text{int}(e_{t\_corr})\)
\STATE \(\delta_{\text{cores}} \leftarrow |\,e_{t\_corr}|\)
\IF{\(e_{t\_corr} > 0\)}
  \STATE $put\_cores\_idle$(\(\delta_{\text{cores}}, \text{cpu\_cores}\)) \label{algo:slctcoreidle:put-idle}
\ELSIF{\(e_{t\_corr} < 0\)}
  \STATE $put\_cores\_active$(\(\delta_{\text{cores}}, \text{cpu\_cores}\)) \label{algo:slctcoreidle:put-active}
\ENDIF
\end{algorithmic}
\end{algorithm}

Overall, the algorithm estimates the age of each core using a rolling idle duration window and distributes the stress of executing inference tasks in a least-aged-first manner. In return, the aging effects of cores take a prolonged time to appear, i.e., slowing down the CPU aging rate. Our age estimation approach avoid the overhead in calculating an accurate aging value. Since \textit{Task-to-Core Mapping} is executed quite frequently in the cloud environments, a minimum execution overhead ensures reduced latency impact on inference request serving.

\subsection{Selective Core Idling}

In addition to the aging rate reduction of \textit{Task-to-Core Mapping}, we provide a core-level optimization mechanism to halt aging, called \textit{Selective Core Idling}. It contributes to the overall aging reduction in the CPU by dynamically halting core aging, whenever the inference task execution is able to tolerate that.

The main idea behind \textit{Selective Core Idling} is to leverage the unused CPU cores in cloud LLM inference for deep idling (see Section \ref{sec:motivation}). We do that by dynamically adjusting the size of the \textit{working set} of cores, and using the remaining cores for deep idling. The key challenge here is to match the size of the \textit{working set} to the number of running inference tasks. If the \textit{working set} is smaller then inference tasks begin to oversubscribe the CPU, whereas a larger \textit{working set} leave a portion of unused cores in the active state which otherwise would have been utilized for deep idling. To address that, we design an algorithm for \textit{Selective Core Idling}. The main part of our algorithm is a module called a reaction function. The reaction function decides the algorithm's sensitivity on adjusting the size of the \textit{working set}. We periodically execute the \textit{Selective Core Idling} algorithm to adjust the \textit{working set} to match the inference throughput.

Algorithm \ref{alg:adjust_sleeping_cores} outlines the proposed algorithm for \textit{Selective Core Idling}. It takes two inputs: the set of available cores, and the number of inference tasks that are oversubscribing the CPU. Once executed, it adjusts the \textit{working set} by selectively setting the idle states of available cores to either deep idle or active. Firstly, the algorithm process the set of available cores to obtain the number of total cores, number of cores that are in the active state, and the number of inference tasks that are allocated with a dedicated core (line \ref{algo:slctcoreidle:counts}). Using them, the algorithm calculates the number of cores that are currently deep idling (line \ref{algo:slctcoreidle:cslp}), as well as the total number of tasks. Here, we cap the total number of tasks at the total number of CPU cores (line \ref{algo:slctcoreidle:capping}). This is done to obtain a normalized error term ($e_{t\_prd}$), which we are calculating next (line \ref{algo:slctcoreidle:error}). The error term indicates the severity of CPU oversubscription of the inference tasks. We then use the error term as the input to the part of our algorithm which carry out the reaction function (line \ref{algo:slctcoreidle:react-start} to line \ref{algo:slctcoreidle:react-end}). The output of the reaction function is then scaled back (line \ref{algo:slctcoreidle:scale-back}) and used to determine the cores to set either active or deep idle. When putting cores to deep idle, we do that in the order of most aged first. When putting cores to active, we do it in the order of least aged first. That way, we complement even-out core aging of \textit{Task to Core Mapping} when deep idling the cores.

\begin{figure}[t!]
  \centering
  \includesvg[width=0.70\linewidth]{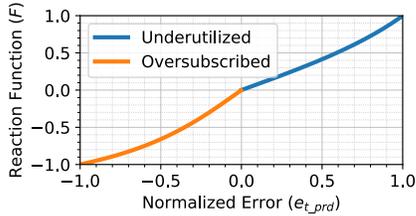}
  \caption{Behavior of the piecewise Reaction Function ($F$) for utilization of the CPU.}
  \Description{Behavior of the piecewise Reaction Function ($F$) for utilization of the CPU.}
  \label{fig:reaction-function}
\end{figure}

\noindent\textbf{Reaction Function}: The reaction function is designed to be independent from the number of CPU cores. It takes the normalized error term ($e_{t\_prd}$) in Algorithm \ref{alg:adjust_sleeping_cores} (line \ref{algo:slctcoreidle:error}) as the input. It returns a normalized output between $-1$ and $+1$. Here a positive output indicates CPU underutilization, requiring to set cores from active to deep idle. Whereas a negative output indicates CPU oversubscription, requiring to set cores from deep idle to active. In the inference cluster, CPU oversubscription impact inference latency of user requests, which is a short term effect requiring immediate action. Whereas CPU underutilization leads to aging of cores, which is a long-term effect since aging is a slow process. To balance this trade-off, we design the reaction function to react slower for CPU underutilization and faster for the CPU oversubscription. Figure \ref{fig:reaction-function} illustrates the behavior of our reaction function. Algorithm \ref{alg:adjust_sleeping_cores} denotes the equation and the values we used for that (line \ref{algo:slctcoreidle:react-start} to line \ref{algo:slctcoreidle:react-end}). 

\section{Implementation}
\label{sec::implementation}

We implement our proposed technique in a simulated environment. We use splitwise-sim \cite{patel24splitwise}, an event-driven, high-fidelity LLM cluster simulator from Microsoft. It employs Splitwise, a state-of-the-art phase splitting LLM serving technique \cite{patel24splitwise}. Compared to vanilla request level scheduling, Splitwise increases CPU load due to the facilitation of additional tasks of phase splitting. As a result, our simulation environment enables exposing our technique to realistic CPU stress levels present in state-of-the-art cloud LLM inference clusters. We extend splitwise-sim to model CPU aging, implement our proposed technique and baselines, and evaluate embodied carbon optimization performance in 2760$^+$ lines of python code. The code for the extended simulator, splitwise-sim-cpu-carbon, is open source at \href{https://github.com/tharindu-b-hewage/splitwise-sim-cpu-carbon}{https://github.com/tharindu-b-hewage/splitwise-sim-cpu-carbon}.

First, we extend the simulator to model the inference tasks that run on the CPU. Table \ref{tab:cpu_tasks_extended_simulator} outlines the tasks we modeled. We model the CPU load of the executor component that facilitates the inference workflow, worker instance tasks that handle memory and iterative-level scheduling, and the tasks of interconnects. For that, we merge each class function with APIs of a new processor subclass we implemented for the CPU. Inside the CPU class, we manage the state of CPU cores. Using the models we outlined in Section \ref{sec::system-model}, we maintain core temperatures, idle states, and the shift in threshold voltage. Each class function call outlined in Table \ref{tab:cpu_tasks_extended_simulator} invoke \verb|assign_core_to_cpu_task| API of the CPU class. It then provide inputs and invoke the Algorithm \ref{algo:task-to-core-mapping}. In return, we determined a CPU core to cater the function call. We update the idle states and temperature of the core to match the task execution. Further, we update the shift in the threshold voltage and the resulting operating frequency of the core. The execution time of the calling function in the simulator is then adjusted according to the operating frequency. In parallel to that, we periodically invoke the \verb|adjust_sleeping_cores| API of the CPU class to conduct \textit{Selective Core Idling}. It retrieve the system state and execute the Algorithm \ref{alg:adjust_sleeping_cores}. In return, the algorithm set the idle state of a number of CPU cores to either active or deep idle. Since the periodic execution of the algorithm \ref{alg:adjust_sleeping_cores} does not add an overhead to inference request latency, we use it as an opportunity to accurately calculate degraded core frequency due to aging. We assume that in practice, that data is provided by the core-level aging sensors \cite{gnad15hyat} with an additional overhead.

\begin{table}[t!]
\centering
\caption{Tasks modeled as inference tasks in the extended splitwise-sim \cite{patel24splitwise} simulator.}
\label{tab:cpu_tasks_extended_simulator}
\begin{tabular}{p{0.3\linewidth}p{0.6\linewidth}}
\toprule
\textbf{Task Name} & \textbf{Class/Function} \\
\midrule
\texttt{finish\_flow}       & \texttt{Executor.finish\_flow}   \\
\texttt{finish\_request}    & \texttt{Executor.finish\_request}\\
\texttt{finish\_task}       & \texttt{Executor.finish\_task}   \\
\texttt{submit}             & \texttt{Executor.submit}         \\
\texttt{submit\_chain}      & \texttt{Executor.submit\_chain}  \\
\texttt{submit\_flow}       & \texttt{Executor.submit\_flow}   \\
\texttt{submit\_task}       & \texttt{Executor.submit\_task}   \\
\texttt{alloc\_memory}      & \texttt{Instance.alloc\_memory}  \\
\texttt{free\_memory}       & \texttt{Instance.free\_memory}   \\
\texttt{start\_iteration}   & \texttt{ORCAInstance.start\_iteration} \\
\texttt{flow\_completion}   & \texttt{Link.flow\_completion}   \\
\bottomrule
\end{tabular}
\end{table}

\section{Performance Evaluation} \label{sec:perf-eval}

In this section, we evaluate the performance of our proposed CPU core management for amortizing embodied carbon in LLM  clusters.  We provide our experimental design and setup, and compare our results with state-of-the-art baselines and analyze them in detail.

\begin{figure*}[htbp!]
    \centering
    \subfloat[Comparison of 40 VM cores]{
        \includesvg[width=0.49\linewidth]{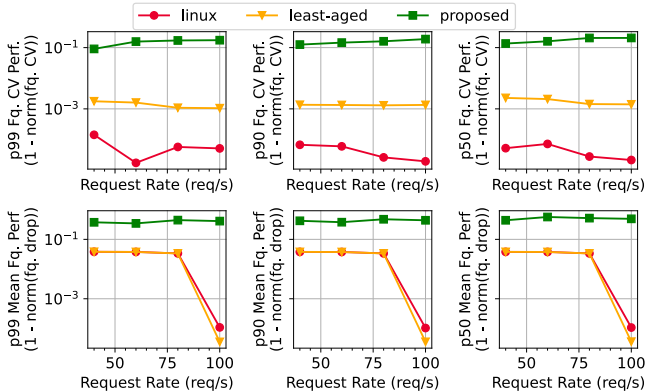}%
        \label{fig:results:aging:vm40-all}
    }%
    \hfil
    \subfloat[Comparison of 80 VM cores]{
        \includesvg[width=0.49\linewidth]{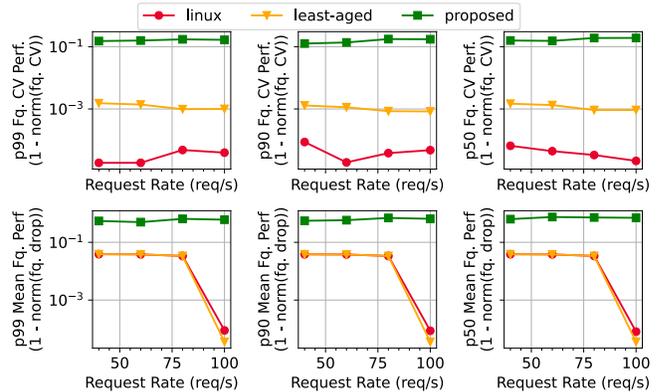}%
        \label{fig:results:aging:vm80-all}
    }%
    \caption{Comparison of managing  aging effects  in CPU.}
    \Description{Comparison of managing  aging effects  in CPU.}
    \label{fig:results:aging}
\end{figure*}

\subsection{Experiment Design and Setup} 

We conduct evaluation experiments in the simulated environment that we modeled and implemented with an LLM cluster simulator from Microsoft. We describe inner details of our implementation in Section \ref{sec::implementation}. We model a cluster of 22 GPU-optimized Nvidia H100 machines with 5 prompt instances and 17 token instances of phase splitting \cite{patel24splitwise}. Our cluster design is an iso-throughput, power-optimized cluster design for cloud LLM inference \cite{patel24splitwise}. We use it to create a realistic cloud inference cluster. Each server in the cluster runs a worker instance. For the worker instance CPU counts, we use CPU core counts of 40 and 80 to match public VM offerings in Azure for Nvidia H100 machines \cite{azure24h100-vms}.

\subsubsection{Baselines:} We compare our technique with the following two baselines:

\noindent  \textbf{\textit{linux: }} It represents executing the servers with the task to CPU core allocation of Linux LLM inference servers. To implement that, we use CPU data from a LLM inference server, captured while executing inference requests \cite{wilkins24llm-energy}. Using the data, we build a probabilistic model to generate inference task to CPU core mappings.

\noindent  \textit{\textbf{least-aged} \cite{zhao2023unsustainableaffinity}: } It is an aging-aware task-serving idea proposed for cloud servers. Unlike most works, \textit{least-aged} proposes the idea of assigning the tasks away from aged cores using executed work as an aging estimate, without requiring frequent CPU profiling. Although \textit{least-aged} was designed for cloud CPU tasks in general, the task characteristics it uses for aging apply to the inference tasks that we model.

\subsubsection{Workloads:} We use LLM inference traces generated by Microsoft using real Azure inference data \cite{patel24splitwise}. Each request in the trace is characterized with the number of input tokens and the number of output tokes generated. It does not provide the actual query that was used in the public cloud environment due to privacy requirements. For our performance metrics that we outline next, the actual query does not make an impact, rather the execution times resulted from processing input and output tokens.

\subsubsection{Metrics:} To measure the CPU aging effects, we use the coefficient of variation (CV) of the distribution of frequencies among CPU cores in each inference server after the experiment. We then calculate the percentile values of that across the cluster. The resulting \textbf{frequency CVs} reflect how well the technique could even out the aging effects among the cores in the cluster machines. To measure the application impact, we calculate the distribution of the number of idle CPU cores in inference servers during the experiments. The resulting data reflect the impact of \textbf{CPU oversubscription} across the cluster servers.

\subsection{Results and Analysis}
\begin{figure*}[htbp!]
  \centering
  \includesvg[width=0.85\linewidth]{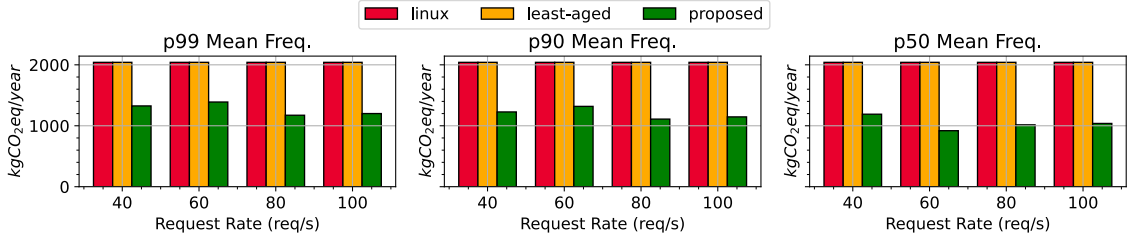}
  \caption{Comparison of estimated yearly CPU embodied carbon reduction in the cluster through management of CPU aging effects.}
  \Description{Comparison of estimated yearly CPU embodied carbon reduction in the cluster through management of CPU aging effects.}
  \label{fig:results:embodied-carb}
\end{figure*}

\begin{figure*}[htbp!]
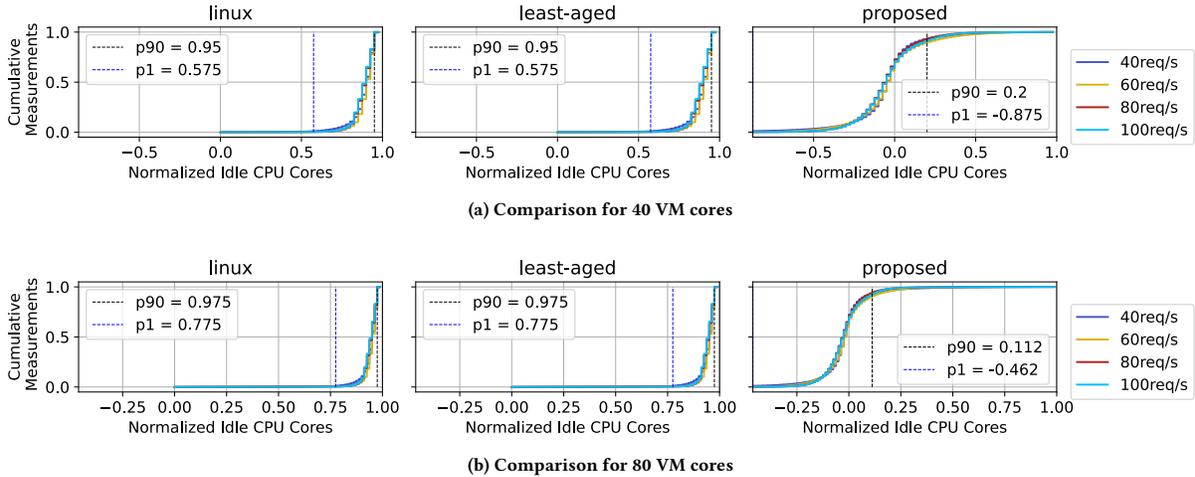

    \centering
    \subfloat[Comparison for 40 VM cores]{
        \includesvg[width=0.9\linewidth]{results/application-impact/vm-cores_40_core_availability_for_task_execution.svg}%
        \label{fig:results:app-impact:vm40}
    }%
    \vfil
    \subfloat[Comparison for 80 VM cores]{
        \includesvg[width=0.9\linewidth]{results/application-impact/vm-cores_80_core_availability_for_task_execution.svg}%
        \label{fig:results:app-impact:vm80}
    }%
    \caption{Comparison of utilization of available cores for running tasks. X-axis in each plot denotes normalized idle CPU cores, where a negative value indicates CPU oversubscription and a positive value indicates CPU underutilization.}
    \Description{Comparison of utilization of available cores for running tasks. X-axis in each plot denotes normalized idle CPU cores, where a negative value indicates CPU oversubscription and a positive value indicates CPU underutilization.}
    \label{fig:results:app-impact}
\end{figure*}

We carry out the performance evaluation as follows. We sample a set of initial core frequencies for each inference server CPU according to the process variation model described in Section \ref{sec:aging-model}. We then replay LLM inference traces on the cluster. We conduct repeated experiments for different throughput levels of LLM inference traces, for each baseline, and for our proposed technique. At the end of the experiments, we calculate the degradation of initial CPU core frequencies through our metrics to evaluate how each baseline and our proposed technique managed cores across the cluster to reduce CPU aging effects and to minimize the application impact. We further estimate the reduction of yearly embodied carbon emissions of the cluster, resulting from the CPU aging management.

\noindent\textbf{Management of core aging effects}: Based on our aging model described in Section \ref{sec:aging-model}, the initial frequency of each CPU core deviates from the nominal value due to process variation. Due to the execution of inference tasks, initial frequencies degrade over time, resulting a frequency distribution across CPU cores. Figure \ref{fig:results:aging} illustrates the results of that. Its subplot, figure \ref{fig:results:aging:vm40-all} illustrates the performance of managing the coefficient of variation (CV) of the core frequency distribution. The performance value decreases when frequency CV increases, and vice versa. It also illustrates the performance of managing the mean frequency degradation. The performance value in that decreases when the mean frequency degradation increases, and vice versa. Both performances are for the VM with 40 cores. Figure \ref{fig:results:aging:vm80-all} illustrates the same for the VM core count of 80. All plots share the x axis, which is the throughput of inference traces.

The results show that in both VM core types, the reduction of frequency variance among cores with \textit{least-aged} is better than the \textit{linux}. It shows the effectiveness of assigning inference tasks away from the aged cores to improve aging imbalance across cores in \textit{least-aged}. However, the proposed technique significantly outperforms both baselines in that. It showcase the superiority of core age even-out behavior across the both \textit{Task to core mapping} and \textit{Selective Core Idling} mechanisms in the proposed technique. The results of managing mean frequency degradation performance shows that both baselines exhibit quite similar performances. In both Figure \ref{fig:results:aging:vm40-all} and \ref{fig:results:aging:vm80-all}, frequency performance values for baselines shows the same for request rates from 40 to 80, and deviates slights at request rate of 100. Whereas our proposed technique consistently surpasses both across all evaluated request rates. It shows the effectiveness of the age halting behavior of the proposed technique. Frequency performance, which is the aging effect of our aging model, shows sustained performance further in-time that the baselines. The discussed performance patterns are consistent across changing VM core sizes.

\noindent\textbf{Reduction of yearly CPU-embodied carbon emissions}: Delayed CPU aging effects allow cloud operators to extend CPU lifespan by increasing the hardware refresh lifecycle. We apply the same with our results of managing core aging effects. We take the hardware refresh cycle of a typical linux-based LLM inference server as 3 years \cite{li24carbon-efficient-llm} and its CPU embodied carbon during this lifespan as 278.3 kgCO\textsubscript{2}eq  \cite{li24carbon-efficient-llm}. We then compare the reduction of the mean core frequency of other techniques to \textit{linux} and estimate an increase in lifecycle extension using a linear model. Using the carbon and lifespan expansion data, we calculate yearly embodied carbon emissions for baselines and the proposed technique. Figure \ref{fig:results:embodied-carb} presents our results. It shows yearly CPU-embodied carbon emissions of the cluster for the age management performance of different throughput levels.

\begin{table*}[htbp!]
  \centering
  \caption{Comparison of Relevant Works with Our Proposed Technique.}
  \label{tab:comparison}
  \renewcommand{\arraystretch}{1.1} %
  \setlength{\tabcolsep}{5pt} %
  \begin{tabular}
  {p{0.12\linewidth}|>
  {\centering\arraybackslash}p{0.18\linewidth}|>{\centering\arraybackslash}p{0.24\linewidth}|>{\centering\arraybackslash}p{0.18\linewidth}|>{\centering\arraybackslash}p{0.18\linewidth}}
    \hline
    \textbf{Work} & 
    \textbf{Even-Out Core Aging} & 
    \textbf{Process Variation Aware} & 
    \textbf{Avoid CPU Profiling} & 
    \textbf{Dynamic Age-halting} \\
    \hline
    Facelift’08 \cite{tiwari08facelift} & 
    \ding{51} & 
    \ding{51} & 
    & 
    \\
 
    Hyat’15 \cite{gnad15hyat} & 
    \ding{51} & 
    \ding{51} & 
     &
    \\
  
    Tamer'21 \cite{saadatmand2021tamer} & 
    \ding{51} & 
    & 
    & 
    \\
    
    Shoulao’23 \cite{Shoulao23} & 
    \ding{51} & 
    \ding{51} & 
    & 
    \\
   
    Zhao’23 \cite{zhao2023unsustainableaffinity} & 
    \ding{51} & 
    & 
    \ding{51}
    & 
     \\
    
    \textbf{Our Proposed} & 
    \ding{51} & 
    \ding{51} & 
    \ding{51} & 
    \ding{51} \\
   \hline
  \end{tabular}
\end{table*}

The results show that yearly CPU-embodied carbon savings with the \textit{least-aged} is minimal when compared to \textit{linux}. This is due to the similar performance of the mean frequency degradation that we discussed previously. In contrast, a cluster managed with our proposed technique shows significant carbon savings in CPU embodied. When estimated with p99 mean frequency performance, our proposed method reduces yearly CPU embodied emissions in our experimental inference cluster by 37.67\%. It further increases to 49.01\% for p50 mean frequency performance. The observed results are due to superiority of mean frequency performance of our proposed technique. It showcase achieving CPU embodied carbon reduction through effective age management. Note that advantage for carbon reductions with \textit{least-aged} over \textit{linux} may improve with the experiment duration. However, goal of our experiments is to evaluate advantage of our proposed technique, which we show within the evaluated experiments.

\noindent\textbf{Application impact of aging-aware core management}: Figure \ref{fig:results:app-impact} illustrates the results of idle cores availability in the cluster servers during inference task execution. The x-axis in all figures shows normalized idle CPU cores, in which a positive value indicates core underutilization and a negative value indicates core oversubscription, whereas, y-axis to denote the measurement distribution.

\par The results show that both baselines do not incur core oversubscription but underutilize cores, yielding positive values of idle CPU cores. p1 to p90 percentiles in both baselines reside closer to 1.0, with a higher VM count increasing the closeness. In contrast, the proposed technique outperforms both baselines in CPU underutilization. Its p90 percentile is at least 77.8\% better in both VM core counts. However, its p1 percentile being negative indicates that the proposed technique does result in CPU oversubscription. The severity of that improves with the VM core count, showing a smaller P1 value. We observe consistent idle core distributions across different inference throughout rates. Additionally, results show that p1 of the proposed technique is at least less than -0.1, which means the proposed technique maintain the CPU oversubscription below 10\%.

In summary, we observe core aging even-out behavior of our proposed technique surpassing baselines in reducing frequency CVs. Alongside, age halting behavior in our technique showcase its superiority in delaying mean frequency degradation. Both frequency CV and mean frequency performance are metrics of CPU aging effects in our system model. We then estimate yearly CPU embodied emissions in the experimental inference cluster, based on the mean frequency performance. Results highlight efficacy of our proposed method to reduce CPU embodied through managing its aging effects. In return, our proposed method show CPU oversubscription, which can impact service quality of the inference tasks. Yet, results show that our proposed technique is able to maintain its severity. 

\section{Related Work}

The environmental impact of embodied carbon in growing LLM inference clusters has caught attention in recent years \cite{bianchini2024dc-power-past-present-future,faiz2024llmcarbon,crusoe24carbon-impact-llm,li24carbon-efficient-llm,nguyen2024sus-llm}. As an early research area, these works model embodied carbon in LLM inference and advocate for potential directions to reduce that \cite{nguyen2024sus-llm, li24carbon-efficient-llm, zhang24emb-crb-models}. Further, some outline CPU GPU asymmetric optimization opportunities of heterogeneous energy, performance, and inference application patterns \cite{li24carbon-efficient-llm, zhang24emb-crb-models}, and accounting carbon footprint for a given inference request on specific hardware settings \cite{fu2024llmco2}. In contrast, to actively optimize embodied carbon, some works propose system-level techniques. These include controlling LLM token generation \cite{li-etal-2024-sprout} and disaggregation of specific compute onto older hardware \cite{shi2024greenllm}. Building on studies of embodied carbon optimization in CPU GPU asymmetric lifetimes, we explore system-level solutions for fine-grain CPU-aging management by leveraging request-level patterns in cloud LLM inference clusters.

A plethora of works investigate mitigating CPU aging effects through workload management \cite{tiwari08facelift,gnad15hyat,saadatmand2021tamer,Shoulao23,zhao2023unsustainableaffinity}. Their predominant approach is even-outing tasks across the cores to reduce uneven aging. These include utilizing CPU profiling \cite{tiwari08facelift,gnad15hyat,saadatmand2021tamer,Shoulao23}, and addressing manufacturing process variations in CPU \cite{tiwari08facelift,gnad15hyat,Shoulao23}. However, not many consider the efficacy of the proposed techniques in cloud settings. For example, conducting CPU profiling in clouds with large server fleets is difficult.
Nevertheless, few recent works consider cloud-efficient techniques, such as workload management at the resource management level to reduce severe exercising of specific cores \cite{zhao2023unsustainableaffinity}. In addition to even-outing aging, age halting is an efficient approach to slow down CPU aging. Age halting has been used in the literature to leverage dark silicon in CPU for age management \cite{gnad15hyat}. However, their age-halting is static since the age-halting adjustments are only done after a relatively longer epoch. In contrast, we study age management for cloud LLM inference with even-outing aging and dynamic age-halting \cite{yahya22agilewatts}.

\section{Conclusions and Future Work}

Given the growing imperative for sustainable growth of cloud LLM inference clusters, our work focuses on mitigating the accumulation of embodied carbon that mostly concentrates on inference server CPU. We propose an aging-aware CPU core management technique, showcasing its potential to extend the cluster's embodied carbon amortization through increasing CPU lifespan. Exploiting CPU underutilization patterns that we uncover, the proposed technique not only even-out silicon aging across cores but also harnesses the opportunities of age halting using core deep idling. Our empirical simulations demonstrate the superiority of the proposed technique over existing methods in reducing the cluster's yearly embodied emissions with a minimum impact on the inference service quality. Our technique enables LLM inference to reduce embodied carbon through the CPU and improve performance with the GPU via the CPU GPU's asymmetric lifetime.

In future work, we plan to implement the proposed technique with cloud resource management middleware and leverage runtime core telemetry data to improve the core aging estimation.

\bibliographystyle{ACM-Reference-Format}
\bibliography{manuscript}

\end{document}